\documentclass{elsart}
\usepackage{epsfig}
\newcommand{\Epa}{\epsilon_{p_a}}

\begin{document}

\begin{frontmatter}

\title{Quantum-tail effect in low energy d+d reaction 
in deuterated metals\thanksref{miur}}

\author[caUni,caINFN]{M. Coraddu\corauthref{corr}},
\corauth[corr]{Corresponding author.}
\ead{massimo.coraddu@ca.infn.it}
\author[caINFN,caUni]{M. Lissia},
\author[caUni,caINFN]{G. Mezzorani},
\author[troitsk]{Yu.V. Petrushevich},
\author[toPoli,caINFN]{P. Quarati},
\author[troitsk]{A. N. Starostin}.

\address[caUni]{Dipartimento di Fisica, Universit\`{a} 
di Cagliari,  I-09042 Monserrato, Italy}
\address[caINFN]{Istituto Nazionale di Fisica Nucleare,
Cagliari,  I-09042 Monserrato, Italy}
\address[toPoli]{Dipartimento di Fisica, Politecnico di Torino, 
         I-10125 Torino, Italy}
\address[troitsk]{State Research Centre of Russian Federation, 
                  Troitsk Institute for innovation and fusion research,
                  Centre for Theoretical Physics and Computational Mathematics,
         Troitsk 142190 Moscow reg., Russia}

\thanks[miur]{This 
work was partially supported by MIUR (Ministero dell'Istruzione,
del\-l'Uni\-ver\-si\-t\`a e della Ricerca) under MIUR-PRIN-2003 project 
``Theoretical Physics of the Nucleus and the Many-Body Systems''.
}

\date{16 January 2004}

\begin{abstract}
The Bochum experimental enhancement of the d+d fusion rate
in a deuterated metal matrix at low incident energies
is  explained by the quantum broadening of 
the momentum-energy dispersion relation
and consequent  modification of the 
high-momentum tail of the distribution function 
from an exponential to a power-law.
\end{abstract}

\end{frontmatter}

\section{Introduction}
Anomalous enhancement of the sub-barrier nuclear fusion 
reaction d(d,p)t in a deuterated metallic matrix has been 
experimentally observed at energies of the incident beam
lower than few keV \cite{Raiola:02,Raiola:02a}. 
Electron screening is not sufficient to explain this enhancement,
and other quantitative explanations are 
missing~\cite{Fiorentini:2002vj}.

In fact, the calculated (adiabatic limit) electron screening 
potential energy $U_{ad}$ is 28~eV, but experiments
show an exponential enhancement of the cross section at very low 
energies that would correspond to a screening energy
$U_{ex}=309 \pm12$~eV \cite{Raiola:02}. Similar behaviors are
found in gaseous targets.

These discrepancies between the calculated $U_{ad}$ and
the experimentally inferred $U_{ex}$ have not yet been
understood \cite{Fiorentini:2002vj}; these puzzling results
could have important consequences also for the study of 
nuclear fusion in astrophysical 
environments~\cite{Coraddu:1998yb,Coraddu:2000nu}.
In this paper we discuss a possible explanation of this enhancement
which is based on the quantum-uncertainty dispersive effect 
between energy and momentum that was proposed 
by Galitskij and Yakimets \cite{Ga:67} and recently discussed 
and applied by Starostin et al. \cite{St:02,St:00,Savchenko:1999ap}.
\section{Anomalously large electron screening}
\begin{figure}[hp]
\begin{center}
\epsfig{figure=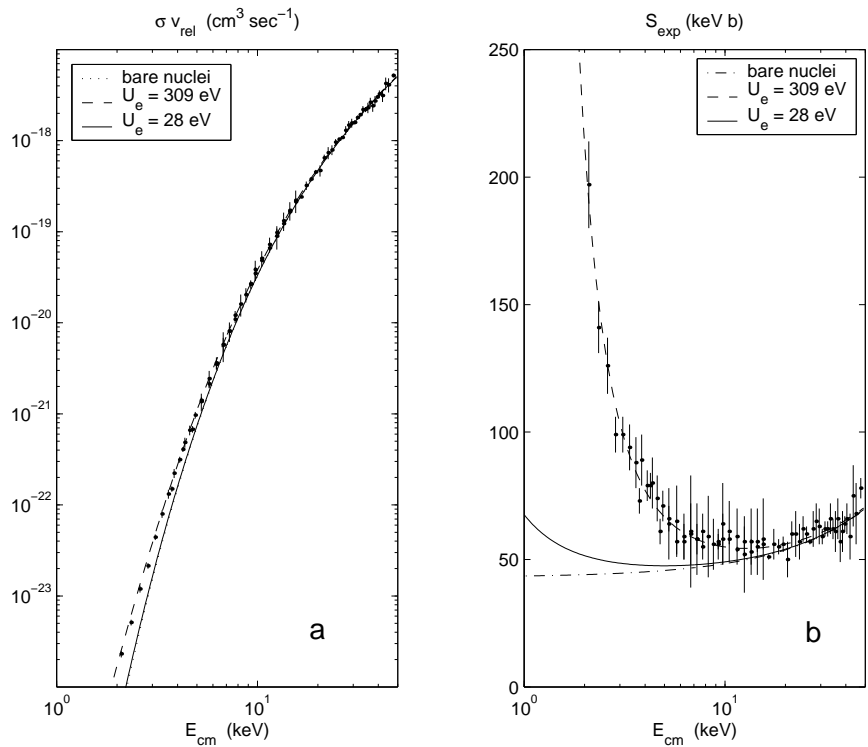,height=10cm}
\caption{Measured values for (a) $\sigma \cdot v_{rel}$
and (b) the astrophysical factor
$S = \sigma(E_{cm}) \cdot E_{cm} \times \exp(\sqrt{E_G/E_{cm}}) $.
Data are compared to the ``bare curve'' of 
Eq.~(\ref{eq:BareCrossSec}) and  the ``screened curve'' 
of Eq.~(\ref{eq:ScreenedCrossSec}) for two values of 
the electron potential: the theoretical adiabatic
upper limit $U_e = 28$~eV
and a fit to the experimental data $U_e = 309$~eV 
\cite{Raiola:02,Raiola:02a}.
\label{fig:RaiolaSigmaV}
}
\end{center}
\end{figure}

Experimental data for the d(d,p)t reaction in a deuterated 
Tantalium target~\cite{Raiola:02} are reported for beam energies
in the range 4-20~keV. 
The target is cooled with nitrogen at a 
temperature of 10 $^{\circ}$C, which corresponds to a thermal energy
$k T = 2.44\times 10^{-2}$~eV. 
Data from Ref.~\cite{Raiola:02} are plotted in 
Fig. \ref{fig:RaiolaSigmaV}.

In panel (a) of Fig. \ref{fig:RaiolaSigmaV}
the experimental values 
of $\sigma(E_{cm}) v_{rel}(E_{cm})$ are plotted versus 
the center of mass energy $E_{cm}$.
Since the incoming particles are not relativistic and the 
target particles are practically at rest, these data
are plotted using 
$ v_{rel}(E_{cm}) = \sqrt{2 E_{cm}/ \mu} $ and 
$E_{cm} = E_{beam} /2 $, where $\mu = m_D/2$ is the reduced mass).

The unscreened cross section can be written as:
\begin{equation}
     \sigma_b(E_{cm}) =
\frac{S(E_{cm})}{E_{cm}}\, \exp{\left( - \pi\, 
   \sqrt{ \frac{E_G}{E_{cm}}} \right)}
     \label{eq:BareCrossSec}
\end{equation}
with $ E_G = 2 \mu \; Z_1^2 Z_2^2 e^4 / \hbar^2 \, $.
The astrophysical factor  $S(E_{cm})$  should vary slowly 
in this energy range and is linearly approximated as 
$S(E_{cm}) =\, S_0\, +\, S_1 E_{cm} \, $, 
where $S_0$\/ and $S_1$\/ are extrapolated from energies  
$E_{cm} > 20$\/ keV (see panel (b) in Fig. \ref{fig:RaiolaSigmaV});
the values reported in Ref.~\cite{Raiola:02} are 
$S_0 = 43$~keV~b and  $S_1 = 0.53$~b.

The interacting nuclei ``feel'' a effective potential barrier
lower by an amount equal the electron screening potential
$U_e$: the resulting screened cross section is:
\[
     \sigma(E_{cm})\, =\, \sigma_b(E_{cm} + U_e) \; .
\]
When  $U_e \ll E_{cm}$\/ a correction factor $f_e$\/ can be factorized:
\begin{equation}
     \sigma(E_{cm})\, \simeq\, f_e \cdot \sigma_b(E_{cm}) \quad ,
     \label{eq:ScreenedCrossSec}
\end{equation}
where
\begin{equation}
     f_e(E_{cm} , U_e) =
\exp{\left( \pi\, \sqrt{\frac{E_G}{E_{cm}}} \cdot \frac{U_e}{2 E_{cm}} \right)}
\quad .
     \label{eq:EScreenFact}
\end{equation}
The electron screening potential computed in the
adiabatic limit $U_{ad}$ constitutes a
theoretical upper limit for $U_e$:
\[
    U_e\, \leq\,U_{ad}\quad .
\]
For  d+d reactions,
Fig.~\ref{fig:RaiolaSigmaV} compares
the ``bare curve'' of Eq.~(\ref{eq:BareCrossSec}) and 
the ``screened curve''  ($U_e = U_{ad} = 28$ eV) of 
Eq.~(\ref{eq:ScreenedCrossSec}) with
the experimental data.
The screened cross section with  $U_e = U_{ad}= 28$ eV
underestimates the experimental data by about
an order of magnitude in the few keV energy range;
the data could be fitted only by using an unphysical
electron potential much larger than the adiabatic upper limit:
$U_e = 309$\/ eV $ \gg U_{ad}$.
\[ 
 S_{exp}=
\sigma(E_{cm})\, E_{cm}\, \exp{\left(\pi\, 
\sqrt{\frac{E_G}{E_{cm}}} \right)}
\quad .
\]
Equation (\ref{eq:ScreenedCrossSec}) 
implies
\[ 
   S_{exp}\, =\, f_e \cdot S(E_{cm})
\]
and, once again, only the screened curve with $U_e = 309$\/ eV 
fits the data.

In summary, there is experimental evidence of anomalously high
values of low-energy fusion cross sections that would require 
electron screening potentials $U_e$ one order of 
magnitude larger than their adiabatic limit,
if explained in terms of screening. 
In deuterated metal targets, the effect depends strongly 
on the metal \cite{Raiola:02a}).
Values of $U_e > U_{ad}$ are required also for describing
experiments with gas target, 
but violations of the requirement $U_e\leq U_{ad} $  are less 
strong than in  metal targets.

These anomalous values of $U_e$ are substantially unexplained;
the screened potential approach is probably trying to mimic 
important processes that have been disregarded.
One should attempt to find alternative
mechanisms that could reproduce the enhancement of the
cross section at low energy. 
 
\section{Thermal corrections and quantum uncertainty }

The parameters of the experiments (temperature and beam energy) are
such that thermal corrections can be neglected. In fact
if we average $\sigma\, v_{rel}$  over a  Maxwellian distribution
of velocity, the temperature correction factor 
$f_T$\/ can be estimated \cite{Fiorentini:2002vj}:
\[
   \langle \sigma v_{rel}  
\rangle_M = f_T \cdot (\sigma v_{rel})_{T=0} \quad ,
\]
\begin{equation}
   f_T = \exp\left( \frac{ \pi^2 E_G\, 
k T}{2\, E_{beam}^2} \right) \quad .
   \label{eq:MaxTermCorrFact}
\end{equation}
For d+d reactions ($\pi^2 E_G = 986$~keV) even at room temperature 
($k T \sim 10^{-5}$~keV) and at the lowest energies of the beam
($E_{beam} \sim 1$ keV) the Maxwellian thermal factor $f_T\simeq 1$.

As showed by  Galitskij and Yakimets \cite{Ga:67} 
many-body collisions broaden the relation between momentum and energy of
the particles.  Since momentum rather than energy  determines
the scattering amplitude, the reaction cross section must be averaged 
over a momentum distribution that may differ from the energy one.
The reaction rate for a  mono-energetic beam is:

\begin{eqnarray}
 \langle \sigma v_{rel} \rangle 
& = & \int d^3\mathbf p_a\, v_{rel}\, \sigma(E_{cm}) f(p_a) \\
   \label{eq:ReacRate}
 f(p_a) 
& = &  \int_0^\infty d E_a  n(E_a)\, \delta_{\gamma}(E_a - \Epa) \quad ,
   \label{eq:MomentDistFunc} 
\end{eqnarray}
where the center of mass energy $E_{cm}$ is function of the
beam-particle and target-particle momentum, and 
the momentum distribution function of the target particles 
$f(p_a)$ depends on  thermal distribution  of the target-particle
energies $ n(E_a)$, which we take to be Maxwellian, and on the 
probability that a target particle with energy 
$E_a$ has momentum $p_a = \sqrt{2 m_a \Epa}$
(dispersion relation), $ \delta_{\gamma}(E_a - \Epa)$.

Quantum effects are responsible for the dispersion relation between
energy and momentum $ \delta_{\gamma}(E_a - \Epa)$ not being a
$\delta$-function, but a broader distribution.
According to  Galitskij and Yakimets \cite{Ga:67} the
relation between energy and momentum is a  Lorentzian 
\begin{equation}
   \delta_{\gamma}(E_a - \Epa) =
 \frac{1}{\pi}\, \frac{\gamma}{\left( E_a - \Epa  \right)^2\, +\, \gamma^2}
   \label{eq:QuantDispRel}
\end{equation}
with $ \gamma = \hbar \nu_{coll} = \hbar\, n\, \sigma_{coll}\, v_{coll} $,
where $\nu_{coll}$\/ and $\sigma_{coll}$\/ are the collisional frequency and 
cross section, while $n$\/ and  $v_{coll}$\/ are the colliding 
particles density and velocity. 

Assuming that the collision cross section could be approximated by
the bare Coulomb cross section 
$\sigma_{coll} =  \pi\, e^4 / \Epa^2 $, the resulting $\gamma$ is
\begin{equation}
   \gamma = 
\frac{\pi\, \hbar N_a \rho\, e^4 }{A\, \Epa^2 }\, 
\sqrt{\frac{2 E_a}{m_a}} 
\quad ,
   \label{eq:WidthQuantDispRel}
\end{equation}
where the relations $v_{coll} = \sqrt{2 E_a / m_a}$ and 
$n = N_A \rho / A$ have been used with $\rho$ the total density 
and $A$  the (average) atomic number.

In the asymptotic regime $ \Epa \gg k T $, relevant 
for the particles that undergo fusion, the Maxwellian
contribution 
$ (2\pi\, m_a\, k T )^{-3/2} \times 
   \exp\left(-\Epa/(k T) \right) $ is negligible
and the distribution of momenta becomes
\begin{equation}
   f(p_a) \sim \frac{1}{(2\pi\, m_a\, k T )^{3/2}} \times
            \frac{\hbar N_A \rho e^4\, k T }{ A\, \Epa^4}\, 
                  \sqrt{\frac{2\pi k T}{m_a}}
\quad .
\end{equation}
Since only particles in the high-energy tail of the distribution
$ \Epa \gg k T $ contribute to the fusion rate, the
quantum effect contribution
\begin{equation}
   \langle \sigma v_{rel} \rangle  \sim 
    \frac{\hbar N_a \rho e^4}{ 2 \pi A\, m_a^2}\, 
 \int   v_{rel}\, \sigma(E_{cm})\, \frac{1}{\Epa^4}\, d^3\mathbf p_a
   \label{eq:QeffTerCorr} 
\end{equation}
is the only important contribution to the rate.

More in general, the fact that the relation between energy and 
momentum is not a $\delta$-function results in a distribution
of momentum $ f(p_a)$ with a power-law asymptotic behavior 
in spite of the energy distribution $n(E_a)$ being exponential.
This power-law tail becomes mostly important for
reactions whose cross sections select high-momentum particles.

This quantum contribution can be calculated numerically and
we are also developing useful analytical approximations: we
shall report elsewhere the derivation of these approximations
and their comparison with the exact numerical integration.

In the following we give some preliminary results using
a parameterization that is 
motivated by the form of Eq.~(\ref{eq:QeffTerCorr}) and that
can be used to qualitatively estimate the
importance of such quantum broadening of the momentum
distribution.

\begin{figure}[hp]
\begin{center}
\epsfig{figure=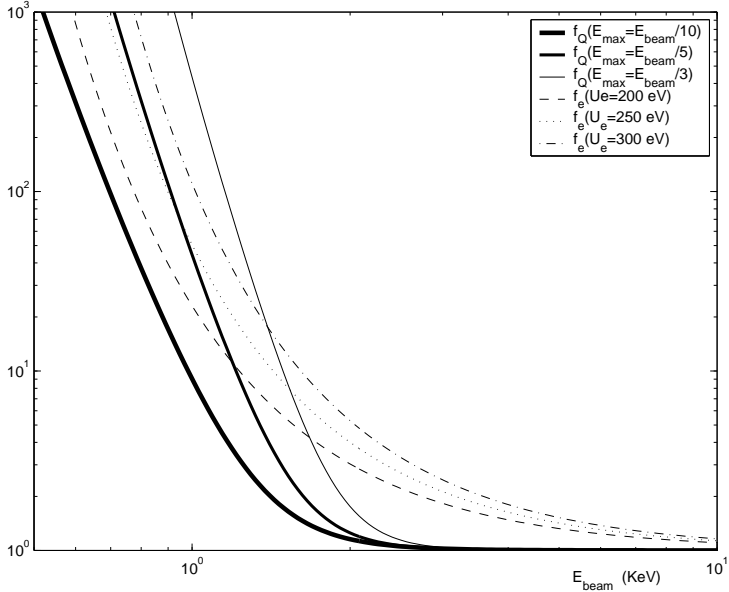,height=8cm}
\caption{The rate 
$f_Q  = \langle \sigma v_{rel} \rangle / (\sigma v_{rel} )_{T=0}$,
where $\langle \sigma v_{rel} \rangle $ is given in 
Eq.~(\ref{eq:QeffLowEnApprox}) for
three values of $E_{max}= E_{beam}/10$, $E_{beam}/5$, and 
$E_{beam}/3$.
For comparison the screening correction factor $f_e$ 
of Eq.~(\ref{eq:EScreenFact}) is also plotted for three 
values of the electron potential $U_e=200$, 250, 300~eV.
\label{fig:EnhancFactCompared}
}
\end{center}
\end{figure}

If we use $ E_{max} < E_{beam}$ as an upper bound 
for $\Epa$ in the integral in Eq.~(\ref{eq:QeffTerCorr})
(the precise value of the low bound is inconsequential)
and
work in the relevant approximation that the target particles
have energies lower than the beam particles, we obtain
the following partial parameterization for the 
dominant contribution
\begin{equation}
    \langle \sigma v_{rel} \rangle   \sim
(\sigma v_{rel})_{T=0}\; \frac{\sqrt{2}}{\pi^2}\, 
           \frac{\hbar N_a \rho\, e^4}{A\, \sqrt{m_a}}\, 
          \frac{E_{beam}^2 
     \exp\left( \frac{\pi \sqrt{2 E_G E_{max}}}{E_{beam}} \right) }
                   {E_G\, E_{max}^{7/2} }\,  \quad .
    \label{eq:QeffLowEnApprox}
\end{equation}

In Fig.~\ref{fig:EnhancFactCompared} we have plotted the ratio
$f_Q  = \langle \sigma v_{rel} \rangle / (\sigma v_{rel} )_{T=0}$    
with   $\langle \sigma v_{rel} \rangle $  given by 
Eq.~(\ref{eq:QeffLowEnApprox}) for three different 
values of $ E_{max}$. 
The function $f_Q$ shows an evident enhancement at low
energy starting from the region of 1-2~keV in 
qualitative agreement with experiments. 
The threshold below which this enhancement becomes 
important depends on $ E_{max} $. We are completing a more
detailed calculation that does not require the introduction
of the parameter $ E_{max}$ and that, 
therefore, can better test the relevance of this quantum
effect for the experimental results.

\section{Conclusion}
The theory of Galitskij and Yakimets predicts that
quantum indeterminacy broadens the relation  between 
energy and momentum.

We have performed a preliminary calculation to estimate the
effect of this broadening on the momentum distribution of 
deuteron in metals and, therefore, on the cross section
of the reaction d(d,p)t. This calculation shows that this
quantum effect should  give an important enhancement of
the cross section at low energies similar to the one
observed in the Bochum experiments.

A more quantitative comparison between theory and
experiments requires the use of more sophisticated analytical
or numerical analyses~\cite{Starostin:2003next}
and the inclusion of the effects of
screening both on the fusion cross section and on the
ion-collision cross section. This further work is being 
completed and will be published in the near future.

\end{document}